\documentclass[fleqn,usenatbib]{mnras}

\usepackage{newtxtext,newtxmath}
\usepackage[T1]{fontenc}
\usepackage{ae,aecompl}

\usepackage{graphicx}	% Including figure files
\usepackage{amsmath}	% Advanced maths commands
\usepackage{amssymb}	% Extra maths symbols

\title[ELBA Survey]{The Environment of Lyman Break Analogues (ELBA) survey: Star-forming galaxies in small groups}

\author[L. Santana-Silva et al.]{
L. Santana-Silva,$^{1}$\thanks{E-mail: luidhy@astro.ufrj.br}
T. S. Gon\c{c}alves$^{1}$,
A. Basu-Zych$^{2, 3, 4}$,
M. Soares-Santos$^{5,6}$,
\newauthor
K. Men{\'e}ndez-Delmestre$^{1}$,
A. Drlica-Wagner$^{5}$,
L. Riguccini$^{1}$,
\newauthor
N. P. Kuropatkin$^{5}$,
B. Yanny$^{5}$,
R. T. Eufrasio$^{7}$
\\
$^{1}$Federal University of Rio de Janeiro, Valongo Observatory, Ladeira Pedro Antonio, 43, Saude 20080-090 Rio de Janeiro, Brazil\\
$^{2}$Department of Astronomy and Center for Space Science and Technology (CRESST), University of Maryland, College Park, MD 20742-2421, USA \\
$^{3}$NASA Goddard Space Flight Center, Greenbelt, MD 20771, USA\\
$^{4}$Department of Physics, University of Maryland Baltimore County, Baltimore, MD 21250, USA \\
$^{5}$Fermi National Accelerator Laboratory, P.O. Box 500, Batavia, IL 60510, USA \\
$^{6}$Department of Physics, Brandeis University, Waltham, MA 02453, USA \\
$^{7}$Department of Physics, University of Arkansas, Fayetteville, AR 72701, USA \\
}

\date{Accepted XXX. Received YYY; in original form ZZZ}

\pubyear{2020}

\begin{document}
\label{firstpage}
\pagerange{\pageref{firstpage}--\pageref{lastpage}}
\maketitle

\begin{abstract}
 The Environment of Lyman Break Analogues (ELBA) survey is an imaging survey of 33 $deg^{2}$ of the southern sky. The survey was observed in {\it u}, {\it g}, {\it r}, and {\it i} bands with the Dark Energy Camera (DECam) on the Blanco telescope.  The main goal of this project is to investigate the
environment of Lyman break analogues (LBAs), low-redshift (z $\sim $0.2) galaxies that are remarkably similar to typical star-forming galaxies at z $\sim$ 3. We explore whether the environment has any influence on the observed properties of these galaxies, providing valuable insight on the formation and evolution of galaxies over cosmic time. Using the Nearest Neighbour method, we measure the local density of each object ranging from small to large scales (clusters of galaxies). Comparing the environment around LBAs with that of the general galaxy population in the field, we conclude that LBAs, on average, populate denser regions at small scales ($\sim$ $1.5Mpc$), but are located in similar environment to other star-forming galaxies at larger scales ($\sim$ $3.0 Mpc$). This offers evidence that nearby encounters such as mergers may influence the star formation activity in LBAs, before infall onto larger galaxy clusters. We interpret this an indication of galaxy preprocessing, in agreement with theoretical expectations for galaxies at z $\sim$ 2 -3 where the gravitational interactions are more intense in early formation processes of this objects

\end{abstract}


%%%%%%%%%%%%%%%%%%%%%%%%%%%%%%%%%%%%%%%%%%%%%%%%%%

%%%%%%%%%%%%%%%%% BODY OF PAPER %%%%%%%%%%%%%%%%%%

%\section{General notes}
%\begin{itemize}
    %\item Figures are referred to as e.g. Fig.~\ref{fig:example_figure}, and tables as e.g. Table~\ref{tab:example_table}.
    %\item Simple mathematics can be inserted into the flow of the text e.g. $2\times3=6$ or $v=220$\,km\,s$^{-1}$, but more complicated expressions should be entered as a numbered equation:

    %\begin{equation}
        %x=\frac{-b\pm\sqrt{b^2-4ac}}{2a}.

	    %\label{eq:quadratic}
    %\end{equation}

%\end{itemize}

\section{Introduction}
Large multi-band galaxy surveys (SDSS-\citealt{york2000}, GALEX- \citealt{martin2005},  2MASS-\citealt{skrutskie2006}, WISE- \citealt{wise}, DES-\citealt{des}) have significantly furthered the efforts of understanding galaxy formation and evolution processes. Based on such studies we know that the galaxy local environment, i.e., the galaxy's immediate surroundings, plays an important role in its formation and evolution, and in driving its observed properties. The first evidence for a correlation between environment and galaxy property was observed by \cite{oemler1974} and \cite{dressler1980}. They showed that in the local universe, spiral galaxies reside in lower-density regions while elliptical galaxies are more likely to be found in high-density environments, the so-called morphology-density relation.
Possible explanations for this empirical relation can be categorized into the two main cases:  processes that occur during the formation of the galaxy (`nature' hypothesis), or those ongoing over the evolution of the galaxy (`nurture' hypothesis). How the latter leads to the morphology-density relation can be explained as follows: in dense environments, mergers and tidal interactions are more common, which could destroy discs in spiral galaxies and convert them into elliptical or lenticular galaxies \citep{toomre1972,farouki1981}.

Many of the processes that act in dense galaxy environments govern the distribution and abundance of gas, which is the fuel for star formation. Therefore, such processes also affect the star formation rate (SFR) within their member galaxies (e.g. ram pressure, strangulation, mergers). Interactions of galaxies in dense intracluster regions can strip the gas, and quench the star formation activity \citep{gunn1972,balogh1998}. Studies have shown that the morphology-density relation is present from low to high density environments \citep{kodama2001,gomez2003}. Several studies have shown that the environment influences galaxy properties, e.g., SFR \citep{kodama2001,gomez2003,lin2014,darvish2016}, galaxy colors \citep{grut2011}, presence of Active Galactic Nuclei \citep[AGN;][]{hatch2014,malavasi2015}, and gas content \citep{darvish2018}. The role of environment on galaxy evolution is complex and benefits from additional studies of environmental density on observed galaxy properties.

Current cosmological models of galaxy formation and evolution describe galaxies in the local universe yet fail to explain properties of galaxies at high redshift \citep[$z=2$--3]{baugh2006}. More complex scenarios than the hierarchical model are required to explain the observed properties in these galaxies, such as cold gas flows and galaxy mergers.  
Lyman break galaxies (LBG) offered a means to select in bulk typical star-forming galaxies at high redshift. Much of our knowledge about the distant ($z>2$) universe comes from their study \cite[e.g. LBGs][]{steidel1999,pettini2001,giavalisco2002,ouchi2004,lee2013,antara2013}.Using cross-correlation methods, \cite{adelberger2005} found that only a small fraction of the baryons in LBG halos are associated with the LBGs themselves. The excess baryons could be associated to other (fainter) objects located in the same halo, that remained undetected below the
LBG survey sensitivity limits.

In our study, we focus on galaxies that share the same physical properties as LBGs, but are low-redshift galaxies and therefore may be subject to different local conditions. \cite{antara2009} studied these low-redshift analogues statistically at $z\sim1$ and concluded that they are mostly found in pairs or small groups. Drawing upon this work, we embark on a study of the local environments of $z<0.3$ analogues using the deep imaging capability of the Dark Energy Camera (DECam, \citealt{decam}). We observed 11 LBAs (Lyman break analogues) that constitute the Environment of Lyman Break Analogues survey.

In this paper, we aim to establish the impact of the environment in the formation of LBAs. The paper is organized as follows: We present our LBA sample in Section 2 and our survey observations in Section 3. Sections 4 and 5 describe our methodology for determining photometric redshifts and the local density of LBAs, respectively. We discuss our results in Section 6 and state our conclusions in Section 7.

Throughout this work, we assume a flat $\Lambda$CDM cosmology with $H_{o} =$ 70 $kms^{-1}$$Mpc^{-1}$ ,$\Omega_{m} = 0.3$, and $\Omega_{\Lambda }= 0.7$. SFRs and stellar masses are based on a Chabrier \citep{chabrier2003} initial mass function (IMF).

%%%%%%%%%%%%%%%%%%%%%%%%%%%%%%%%%%%%%%%%%%%%%%%%%%%%%%%%%%%%%%%%%%
\section{The Lyman Break Analogue Sample}
%%%%%%%%%%%%%%%%%%%%%%%%%%%%%%%%%%%%%%%%%%%%%%%%%%%%%%%%%%%%%%%%%%

Lyman break analogs (LBAs) are a sample of low-redshift galaxies ($z\sim 0.3$) that were selected to reproduce the main properties \cite[e.g.][]{heckman2005} of star-forming galaxies in the distant Universe, i.e., the LBGs, which have served as a relevant and well-explored population in studying galaxy evolution. The LBAs are a subsample of the Ultraviolet Luminous Galaxies (UVLG) identified by matching the Galaxy Evolution Explorer (GALEX) survey to the  Sloan Digital Sky Survey DR-3 by \citet{heckman2005,hoopes2007}. They identified 215 galaxies with FUV luminosities $L_{FUV}> 10^{10} L_{\odot}$, which corresponds to the typical  rest-frame UV luminosities of Lyman break galaxies (LBG) at $z\sim 3.0$ ($10^{10.8}L_{\odot}$, \citealt{arnouts2005}) and to that of lower-redshift  galaxies ($z$= 1.5 - 2.5) that also have LBG-like spectral types, i.e., the BX and BM galaxies \cite[e.g.][]{adelberger2005}.

LBAs are defined as the subsample of UVLGs with the highest surface brightness ($I_{FUV} > 10^{9} L_{\odot}Kpc^{-2}$), characterized by super-compact morphologies and exhibit metallicities, gas fractions, morphologies and attenuation values similar to those found in main sequence star-forming galaxies at high redshift ($2\leqslant z \leqslant 3$);\citep{hoopes2007,antara2007,antara2009,antara2013,roderik2010,roderik2011,thiago2010,thiago2014}. Thereby, LBAs represent a reliable sample for studying the detailed processes that happen in galaxies in the distant universe because they share similar properties and being at low-z they are closer by and hence do not suffer from cosmic deeming, offering a favourable view of the high-z processes. These galaxies are rare and unique in the local universe. They present a spatial density $\sim$ $10^{-5}$ $Mpc^{-3}$\citep{heckman2005}. 

Our main goal in this work is to evaluate the effects of the environment (if any) on the formation of LBAs. For this we observed the fields around each LBA in our sample in four photometric bands (ugri). We selected 11 LBAs based on the observability from Blanco Telescope — these observations constitute the ELBA survey.These are a subsample of the LBA sample from  \cite{heckman2005,hoopes2007}. The 11 LBAs in our sample are representative of the wider LBA population because they cover typical ranges in stellar mass and SF, as well as physical sizes. This allows us to undertake an environmental study of LBAs without selection effects in the sample.

\begin{figure*}
    \includegraphics[width=\columnwidth]{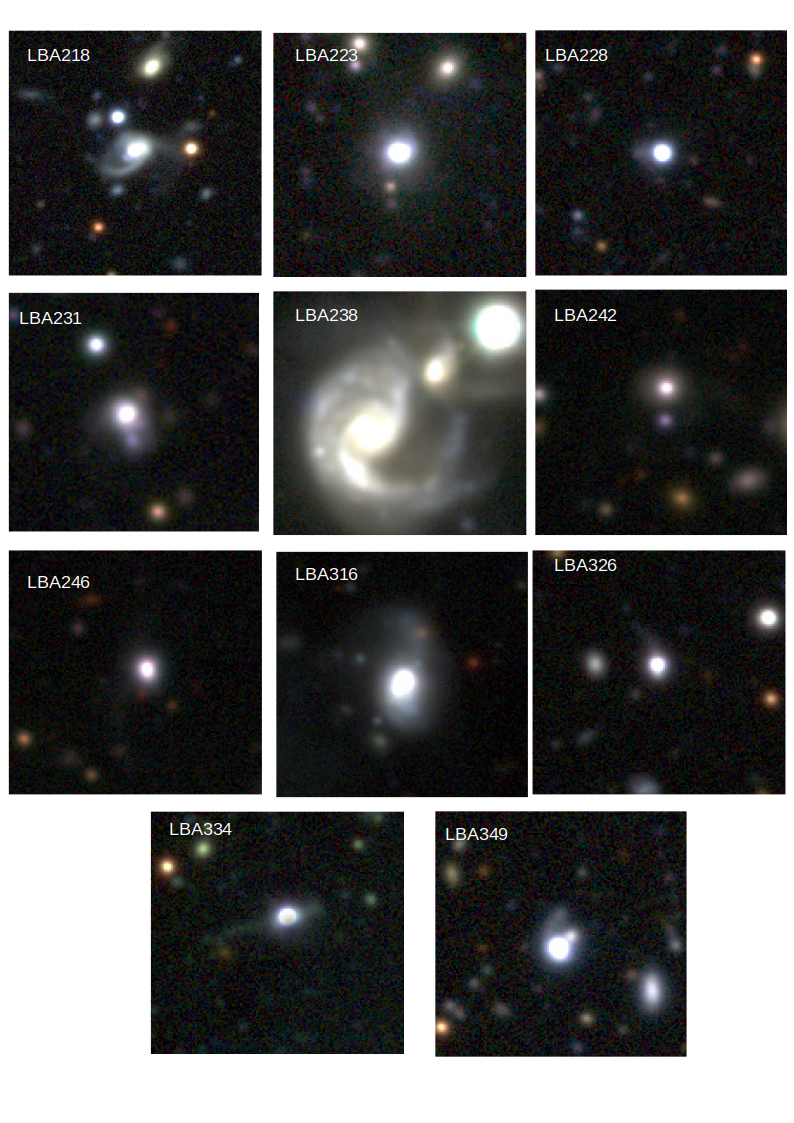}
    \caption{ELBA image cutouts, measuring 25" x 25", produced by a stack of the images observed on {\it g}, {\it r} and {\it i} bands. These images illustrate the compact morphologies of the LBAs. These 25" image cutouts show a range of potential signs from mergers or interactions. For example, a few show nearby companions, while others have faint tidal features (also seen by HST, \citealt{overzier2009}). By encompassing a larger FOV than HST, the ELBA survey probes both the small and the large-scale environments of LBAs.}
    \label{ELBA_sample}
\end{figure*}

\begin{figure}
    \includegraphics[width=0.48\textwidth]{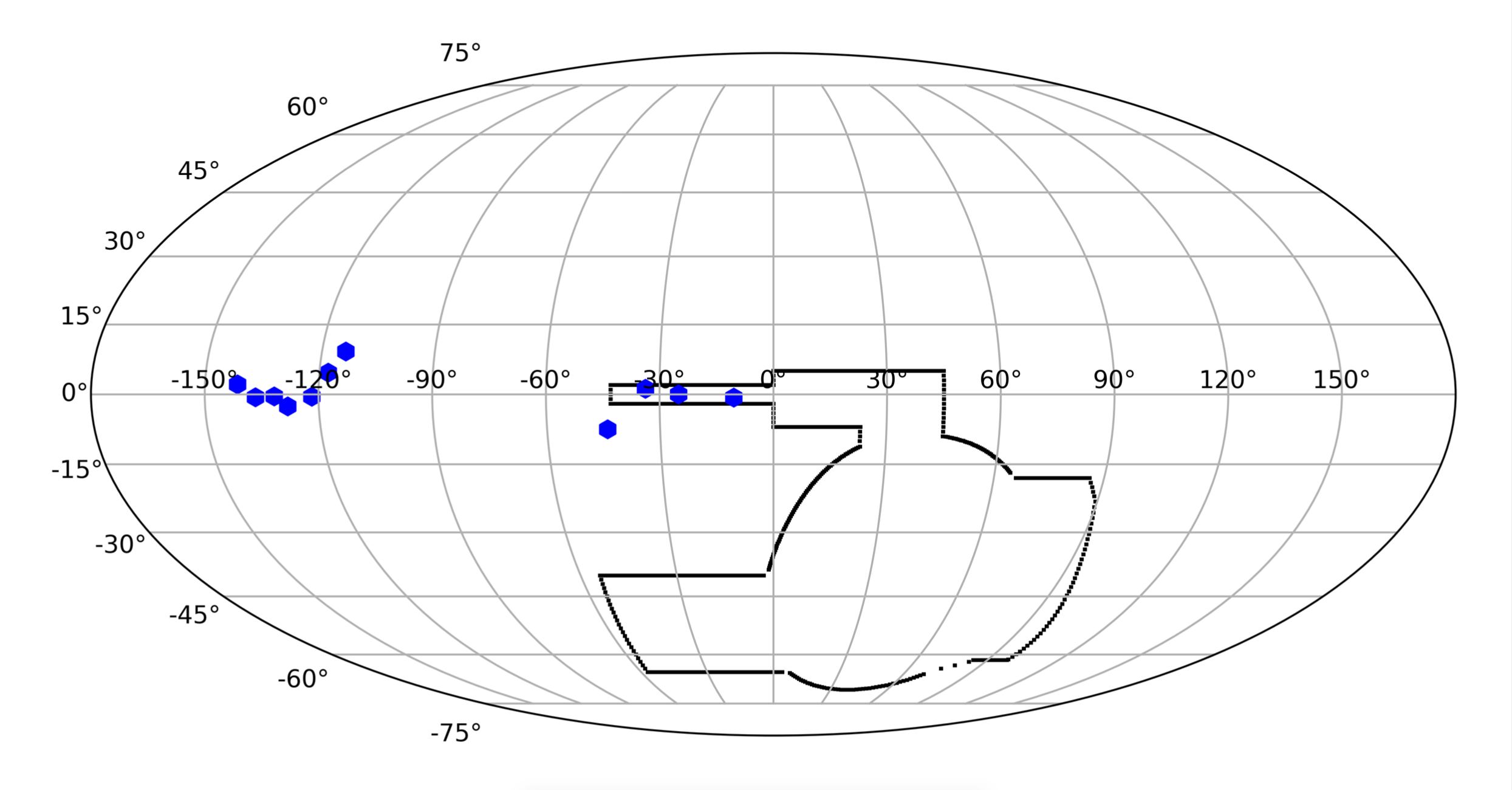}
      \caption{The ELBA survey consists of deep ugri DECam imaging on 11 individual positions (here shown as blue dots), each centered on a Lyman-Break Analogue (LBA). With each field covering a large physical area of 3 square decrees in the sky, we are able to compare the environment of LBAs with that of other galaxies of similar stellar mass or star formation rate.}
    \label{ELBA_sample_sky}
\end{figure}

\begin{table}
    \centering
    \caption{LBAs in the ELBA survey}
    \label{sample}
    \begin{tabular}{c c c c  c c}
    \hline
    \hline
    Galaxy& z  &  $logM_{*}(M_{\odot})$ & $R_{50}$ & SFR$(M_{\odot}yr^{-1})$ \\
    \hline
    LBA218&0.180293&         10.7           &4.61      &7.48    \\
    LBA223&0.185750&         10.3           &2.53      &6.14    \\
    LBA228&0.217992&         9.30           &0.93      &13.16   \\
    LBA242&0.254800&         10.6           &1.57      &7.09     \\
    LBA246&0.221935&         10.0           &2.07      &16.50    \\
    LBA334&0.230648&         9.8            &0.87      &18.99     \\
    LBA315&0.136773&         10.8           &2.77      &23.14      \\
    LBA349&0.251667&         10.0           &3.07      &16.07     \\
    LBA326&0.204322&         10.0           &1.30      &9.18    \\
    LBA231&0.139000&         10.4           &2.46      &3.00    \\
    LBA238&0.079319&         9.8            &0.21      &2.77    \\
    \hline
    \hline
    \end{tabular}
\end{table}

%%%%%%%%%%%%%%%%%%%%%%%%%%%%%%%%%%%%%%%%%%%%%%%%%%%%%%%%%%%%%%%%%%
\section{ELBA Survey}
%%%%%%%%%%%%%%%%%%%%%%%%%%%%%%%%%%%%%%%%%%%%%%%%%%%%%%%%%%%%%%%%%%

The observations were performed between 2014 and 2017 for 10 nights (programs 2014A-0632, 2015A-0619, 2017A-0913) on the Blanco Telescope located at Cerro Tololo Inter-American Observatory (CTIO), using the Dark Energy Camera (DECam). DECam is a mosaic CCD camera composed of 62 CCDs with a field of view of 3 square degrees \cite[e.g.][]{decam} and a plate scale of 0.236 arcsec per pixel. Images were obtained in four broadband filters {\it u},{\it g},{\it r} and {\it i}. The {\it u} band observations were included to obtain photometric data on either side of the 4000 $\AA$ break, yielding more reliable photometric redshifts for the galaxies in the imaged fields. To obtain deep photometry with high signal-to-noise ratio the total exposure times are 3600s for the {\it g} and {\it r} bands, 1800s for the {\it i}-band and 5000s for the {\it u} band. These allow us to obtain reliable 10$\sigma$ detections, down to {\it u} $=$ 22.5 $mag_{AB}$, {\it g} $=$ 24.4 $mag_{AB}$, {\it r} $=$ 24.0 $mag_{AB}$, {\it i} $=$ 23.2 $mag_{AB}$.

We present the physical properties and image cutouts for each of the 11 LBAs in Table \ref{sample} and Figure \ref{ELBA_sample}. The table includes spectroscopic redshifts from Sloan Digital Sky Survey (SDSS), stellar masses, star formation rates and half-light radii ($R_{50}$ in {\it u} band from  \citet{hoopes2007}. In Figure \ref{ELBA_sample}, we show 25" $\times$ 25" image cutouts, centered on each of the  LBAs, and produced by stacking the {\it gri} bands. In this figure we show the 11 galaxies that constitutes our sample. Two galaxies in our sample LBA228 and LBA316 were imaged under poor observational conditions. Throughout this paper our results are based on the nine galaxies with good data quality.

In total, the ELBA survey covers approximately 33 deg$^{2}$ with individual exposures of 100s in {\it ugri}. The complete survey is composed of 11 fields, each field centered around one LBA. The total areal coverage is presented in Figure \ref{ELBA_sample_sky}. Since all observed fields are within the SDSS \cite[e.g.][]{york2000} footprint, a significant fraction of the brightest sources in our data also have SDSS spectroscopic data allowing us to test the quality of our ELBA-derived photometric redshifts.

%%%%%%%%%%%%%
\subsection{Photometry and catalogs}
%%%%%%%%%%%%%%%%%%%%%%%%%%%%%%%%%%%%%%%%%%%%%%%%%%%%%%%%%%%%%%%%%%
In this section, we describe the details in the analysis, including image processing and photometry for detected sources in the data used in this paper.
We coadd images according to the method proposed by \cite{morganson2018}. All single epoch images were processed using DESDM (Dark Energy Survey Data Management). We use SCAMP \citep{scamp} to calculate astrometric solutions to guarantee that the sources will be in the same region or pixel in the final coadded image. To build the image we use SWARP \citep{bertin2002} to resample the individual exposures and build the final image for each field in the observed bands.

We use Source Extractor \citep{bertin1996} to produce the catalogs from coadded images. We use the r-band data as detection images to identify the location for each source in the images. We run SExtractor in double image mode to constrain source positions in each coadded (the detection catalogs) and single image in order to determine source magnitudes in each observed band. The basic parameters available for each object in the catalogs include Flags, position values (ALPHAWINJ2000, DELTAWINJ2000) and the photometry for each observed band (MAGAUTO, MAGPSF). The same methodology was used to reduce and produce source catalogs for the Blanco Imaging the Southern Sky Survey (BLISS) \cite[e.g.][]{sid2019}.

We detect faint galaxies down to $r\sim 25$  with S/N $\sim$ 3. To estimate the magnitude limit in ELBA data, we measure the S/N  using the uncertainties in magnitudes obtained using SExtractor. The magnitude errors in the objects are estimated as the inverse of the S/N corrected by Pogson's constant.

\begin{figure*}
  \centering
  \includegraphics[width=0.73\textwidth]{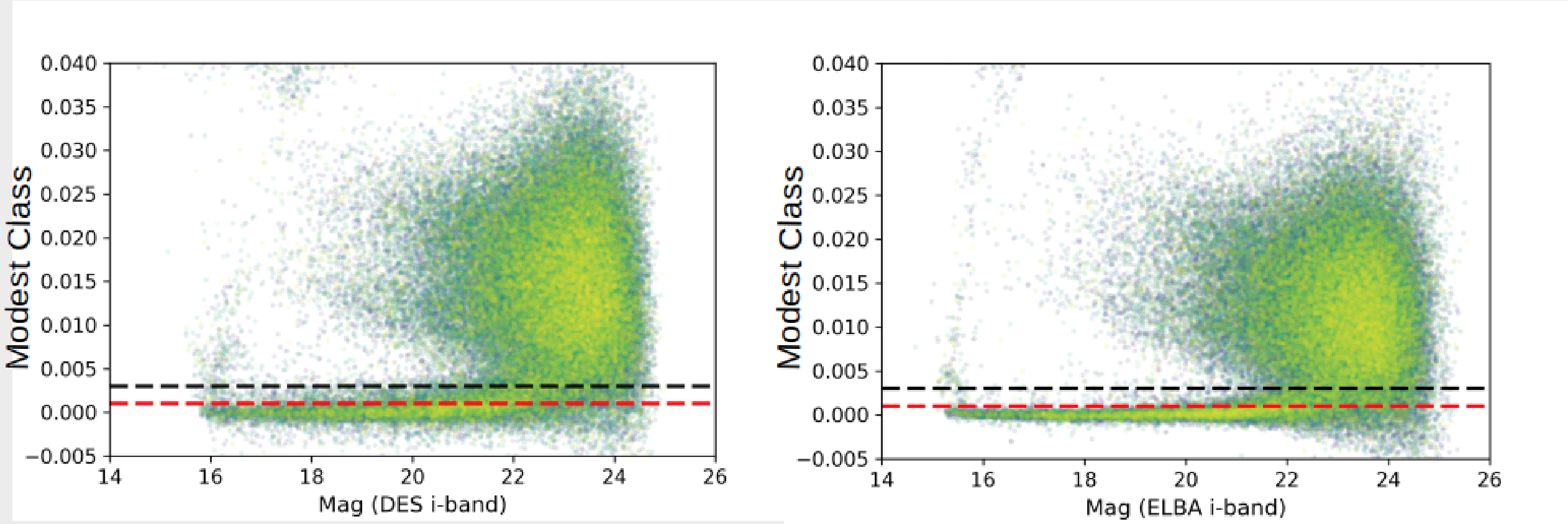}
  \caption{In order to distinguish between extendend and point sources, we use the $Modest Class$ parameter. We fixed this parameter at 0.002, wich means that all sources with modest class values smaller than 0.002 are classified as point sources. In these plots we show the $Modest Class$ selection for the center square degree region around the LBAs analyzed in this work that have match with DES catalogs, with the distribution of the $Modest Class$ as a function of the DES i-band and ELBA i-band shown in left and right panels, respectively. The right panel shows the distribution of modest class as function of ELBA i-band. Both panels clearly identify a clear stellar locus where the $Modest Class$ is close to zero. In all panels the black (red) lines corresponds to thresholds applied to modest class for a pure galaxy selection.The red line represents $Modest Class$ equals 0.002. We use the same parameters as the DES in order to demonstrate that we can use these thresholds on ELBA data with no negative impact in our star galaxy separation.}
  \label{modestclass}
\end{figure*}

\begin{figure*}
    \includegraphics[scale = 0.7]{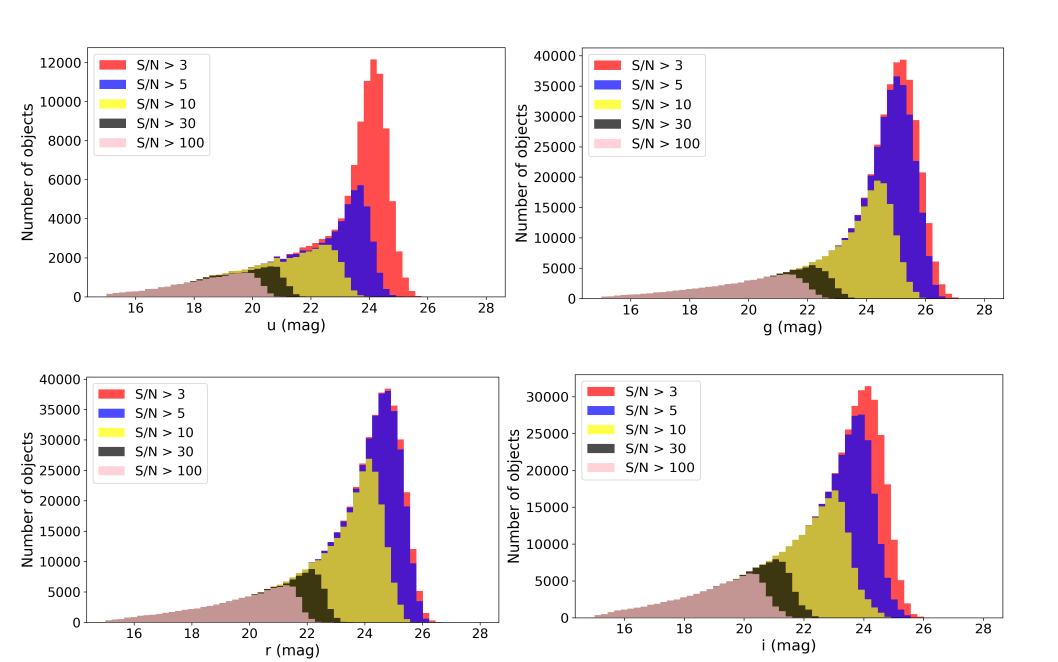}
    \caption{Magnitude distributions for point sources as a function of signal-to-noise ratio for all observed bands on ELBA survey. We adopted the peak of the distribution for sources with S/N $\sim$ 10 as the magnitude limit in each band for the survey.}
    \label{ELBAsurveysn}
\end{figure*}

To perform Star-Galaxy separation we apply the modest class classification used by \cite{alex2018} on DES data. The modest class classification is associated with SPREAD-MODEL and SPREADERR-MODEL in i-band (Figure \ref{modestclass}) and are represented by the expression $SPREADMODEL + 5/3 \times SPREADERRMODEL$. The spread model is a morphological parameter obtained using Sextractor by normalization between the best PSF model and an extended model of a circular disk convolved with the PSF. We adopted $Modest Class \leqslant 0.002 $ for point sources \cite[e.g.][]{alex2018}. The magnitude distribution for point sources as a function of the S/N is presented in Figure \ref{ELBAsurveysn}. In Figure \ref{modestclass} we compare how the star-galaxy separation criterium performs for pure sources from DES and for pure sources on ELBA: the classifications agree with each other. We adopt the peaks of the distributions of magnitudes as the magnitude limit for point sources for a 10 $\sigma$ detection. We obtain 22.5, 24.4, 24.0, 23.2 for {\it u} {\it g}, {\it r} and {\it i} band, respectively.Table \ref{SDSSxELBA} shows a comparison between the ELBA magnitude limit and Dark Energy Survey DR-1 \citep{des}.

\begin{figure}
    \includegraphics[width=\columnwidth]{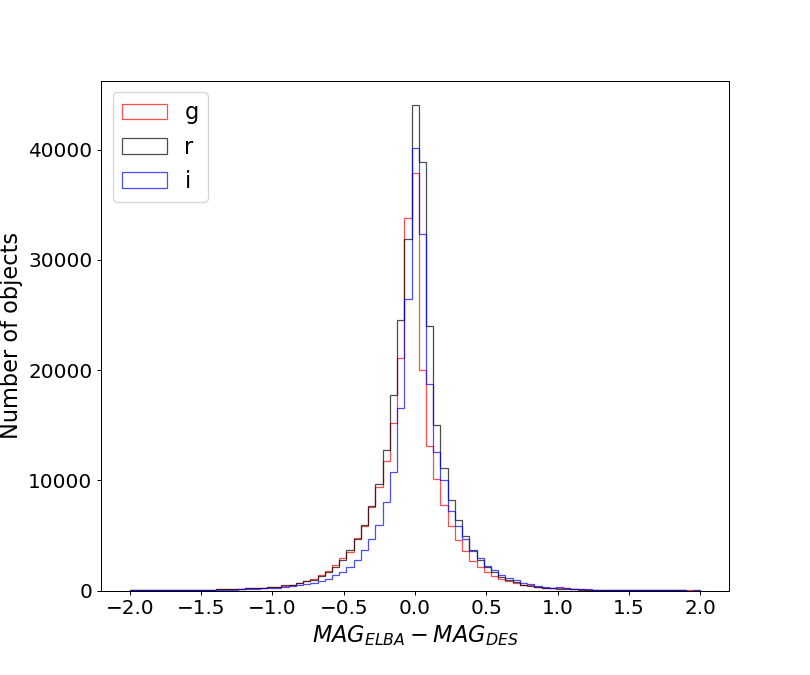}
    \caption{Photometry comparison between ELBA and DES for filters {\it g},{\it r} and {\it i}. In order to test our photometry accuracy we compare our measured magnitudes to those on the DES survey catalog. The difference in magnitude is typically very close to zero, with a difference of 0.02 dex for sources with $S/N \geq 3$.}
    \label{ELBAsurvey2comp}
\end{figure}

The reliability of ELBA photometry was tested comparing the magnitudes of sources in ELBA fields that are available in the DES database (Figure \ref{ELBAsurvey2comp}). The scattering in magnitude for between our data and DES is $\sim 0.01$, The large differences for some sources is due to inaccurate photometry for objects with low surface brightness.

\begin{table}
    \centering
    \caption{Comparison of magnitude limit between ELBA,SDSS and DES for $10\sigma$ detections.}
    \label{SDSSxELBA}
    \begin{tabular}{c c c c c }
    \hline
    \hline
    Filter & ELBA ($10\sigma$) & DES ($10\sigma$) & SDSS ($10\sigma$)\\
    {\it u}& 22.5             & -----             & 22.0                                     \\
    {\it g}& 24.4             & 23.5              & 22.2                                                 \\
    {\it r}& 24.0             & 23.1              & 22.2                                                   \\
    {\it i}& 23.2             & 22.5              & 21.3                                               \\

    \hline
    \hline
    \end{tabular}
\end{table}

\section{Photometric Redshifts}
%%%%%%%%%%%%%%%%%%%%%%%%%%%%%%%%%%%%%%%%%%%%%%%%%%%%%%%%%%%%%%%%%%%
We use LePhare to estimate the photometric redshifts. LePhare \citep{ilbert2006} is a public code that uses template fitting and $\chi^{2}$ minimization between observed and theoretical magnitudes to estimate the best template and redshift. The code provides several template sets and extinction laws already available for the user. For this study we choose the set of templates CFHTLS by \cite{ilbert2006} obtained from the interpolation of the commonly used  \cite{coleman1980} templates with different Hubble types and \cite{kinney1996} for starburst galaxies.

\begin{figure}
  \centering
  \includegraphics[width=0.45\textwidth]{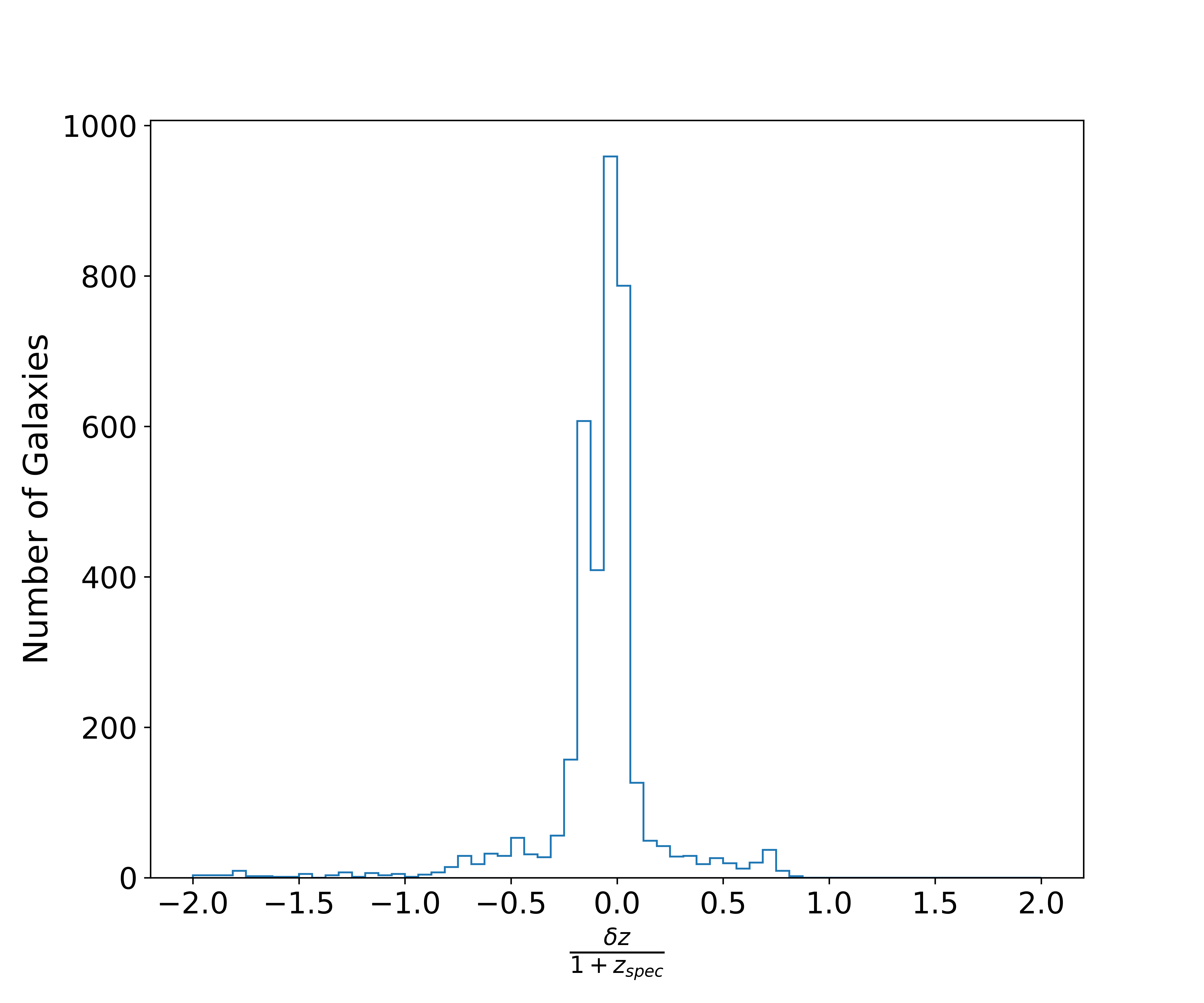}
  \includegraphics[width=0.46\textwidth]{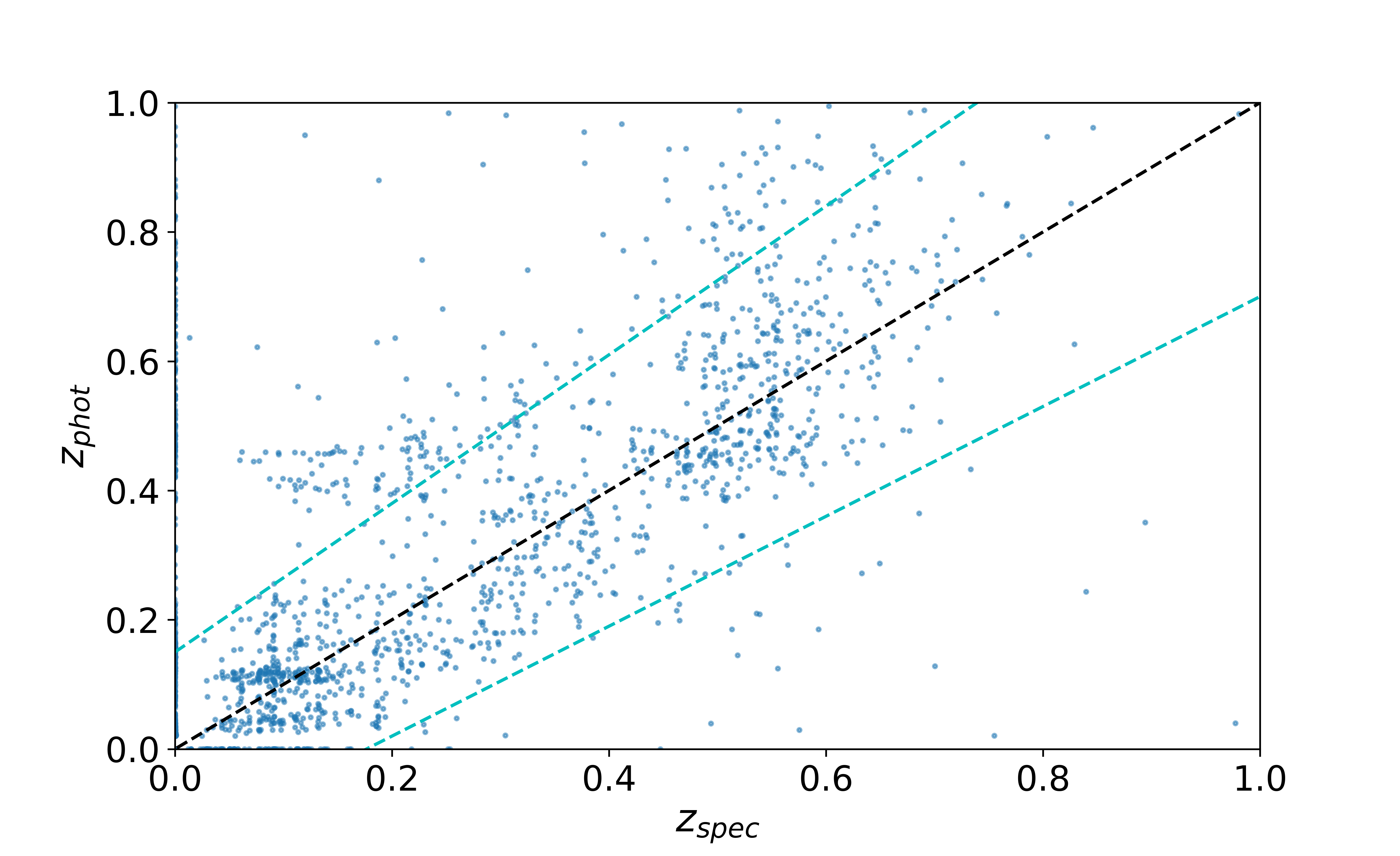}
  \caption{We gauge our photometric redshift accuracy by comparing our ELBA photometric redshifts with spectroscopic redshifts available on Sloan Digital Sky Survey (SDSS). The top panel show the distribution of values for $\delta _{z}/(1+z_{spec})$, for photometric redshift measurements using only the {\it ugri} bands observed on ELBA. The bottom panel shows the correlation between ELBA photometric redshifts and SDSS spectroscopic redshifts. The solid line corresponds to values where the photometric and spectroscopic redshifts are identical and the dotted lines represent  $z_{phot} = z_{spec}\pm 0.15(1+ z_{spec})$. Using only the four bands available on ELBA survey, we recover 75$\%$ of the spectroscopic redshifts at $I_{AB}\leq 24$ with $\sigma _{\delta z/(1+z)} = 0.0684 $ and a median bias of ${\delta _{z}/(1+z)} = 0.0462$. }
  \label{photoz_UGRI}
\end{figure}

To estimate the accuracy of photometric redshifts on ELBA data we used the normalized median absolute deviation (NMAD). NMAD is a robust measurement of the precision obtained in a sample of photometric redshifts with spectroscopic counterparts available \citep{brammer2008}. In addition it yields appropriate uncertainties of the photo-z distribution without being affected by catastrophic errors \citep{molino2014}. The NMAD is defined as:
\begin{equation}
    \sigma_{NMAD} = 1.48\times {\rm median} \left\lvert \frac{\delta_{z} - median (\delta_{z})}{1 + z_{s}}  \right\rvert,
	\label{eq:sigma}
\end{equation}
where $\delta_{z} \equiv z_{b} - z_{s}$ is the difference between the photometric redshift obtained using LePhare $z_{b}$ and the spectroscopic redshift $z_{s}$.
The accuracy of the photometric redshifts using only {\it ugri} is presented in Fig \ref{photoz_UGRI}. In this set of observations using only four bands available on ELBA survey, we recover 75$\%$ of the spectroscopic redshifts at $I_{AB}\leq 24$ with $\sigma _{\delta z/(1+z)} = 0.0684 $ and a median bias of ${\delta _{z}/(1+z)} = 0.0462$.

\begin{figure}
  \centering
  \includegraphics[width=0.45\textwidth]{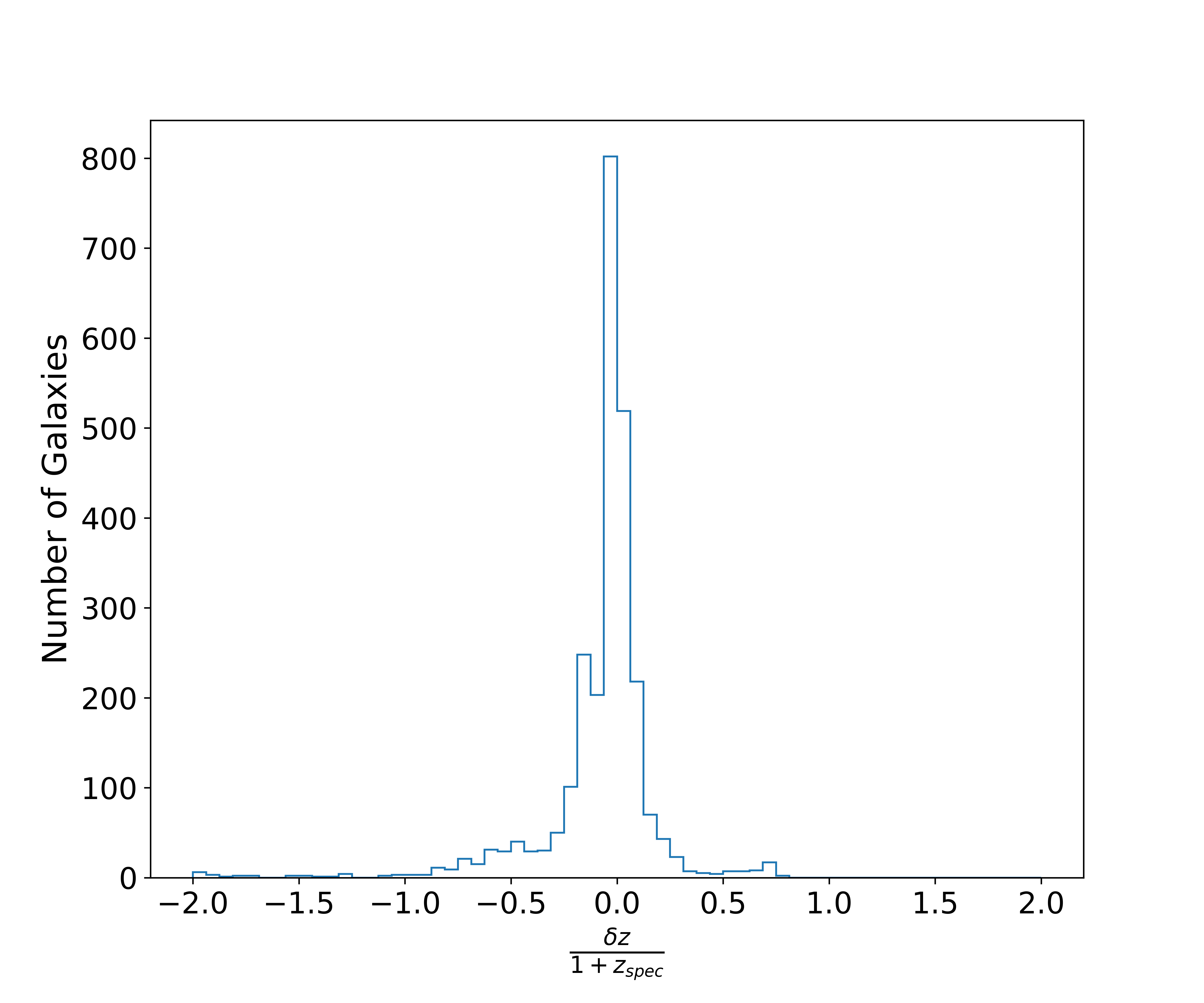}
  \includegraphics[width=0.46\textwidth]{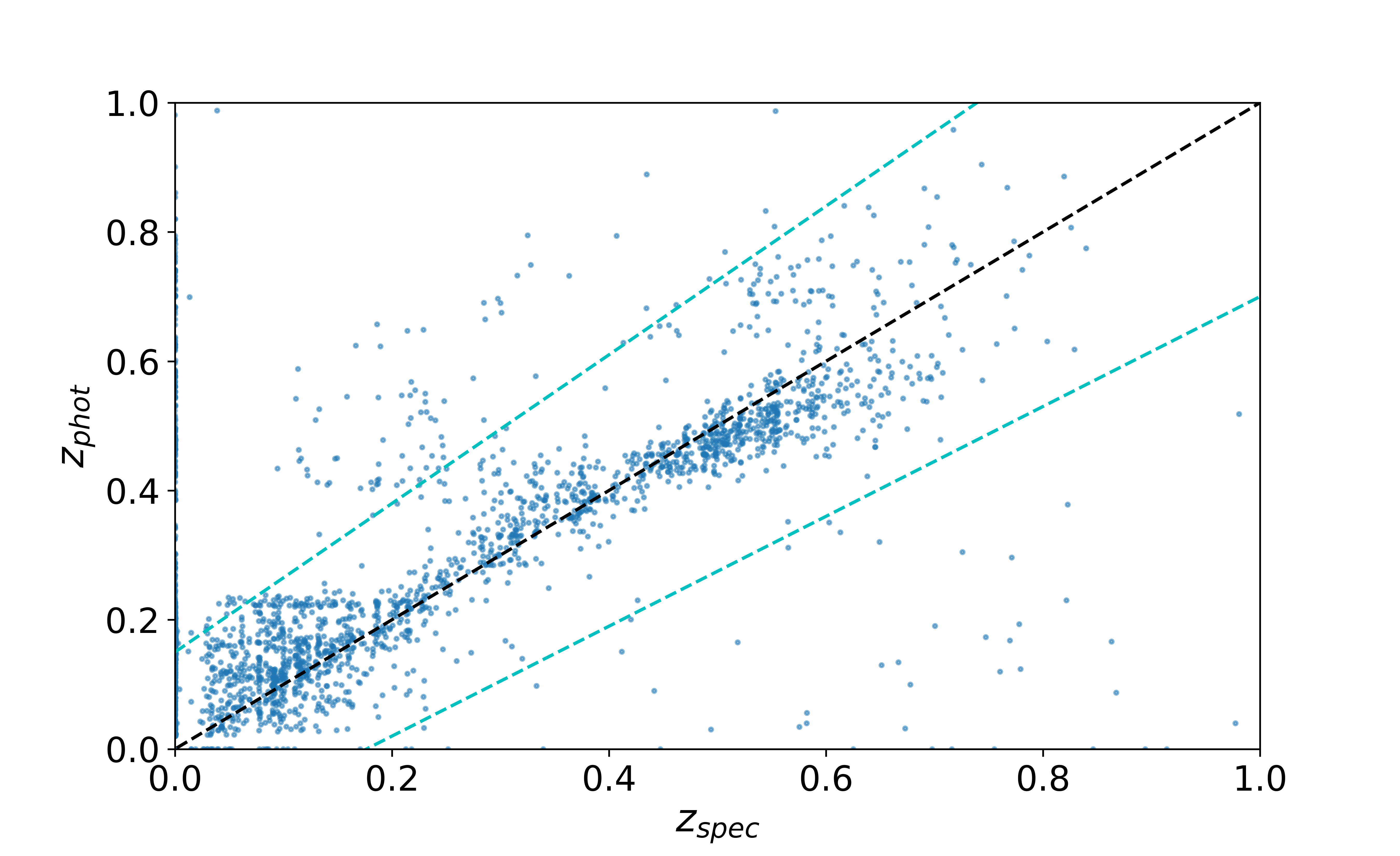}
  \caption{Same as Figure \ref{photoz_UGRI} but now including JHK bands from the UKIDSS survey. Photometric redshifts are vastly improved. Adding the bands JHK available in UKIDSS we perform new photometric redshift measurements and recover 87$\%$ of the spectroscopic redshifts at $I_{AB}\leq 24$ with $\sigma _{\delta z/(1+z)} = 0.0228 $ and a median bias of ${\delta z/(1+z)} = 0.0186$.}
  \label{photoz_UGRIJHK}
\end{figure}

To increase the precision of the photometric redshift measurements, we also include near-infrared photometry from the UKIRT Infrared Deep Sky Survey \citep{ukidss}. The UKIDSS survey covers 7500 degrees over the north sky covering regions in low galactic latitude. Adding the bands JHK available in UKIDSS we perform new photometric redshift measurements and recover 87$\%$ of the spectroscopic redshifts at $I_{AB}\leq 24$ with $\sigma _{\delta z/(1+z)} = 0.0228 $ and a median bias of ${\delta z/(1+z)} = 0.0186$ (Figure \ref{photoz_UGRIJHK}). We use these improved photometric redshifts to perform environmental density measurements.

%%%%%%%%%%%%%
\section{The local density of LBAs}{\label{sec:environmentdensity}
%%%%%%%%%%%%%%%%%%%%%%%%%%%%%%%%%%%%%%%%%%%%%%%%%%%%%%%%%%%%%%%%%%%
In order to measure the local number density around each galaxy (with S/N $\geq$ 2, in order to avoid spurious detections), we use the k-th nearest neighbour (KNN) method. For a given value k we specify the distance to the k-th nearest neighbours to the target galaxy by computing the surface projected density, $\delta_{k}$ , which is defined as:
\begin{equation}
\delta_{k} = \frac{k}{\pi r_{k}^{2}}
\end{equation}
where {\it k} is the number of neighbours and $r_{k}$ is the projected radius distance to the $k$-th nearest neighbor.  However, since we are observing objects in the sky, two galaxies can appear close to each other due to a projection effect while actually being physically distant from each other. One generally adopts velocity cuts or redshift slices around the galaxy target in order to mitigate such projection effects. This cut is typically of order $\pm$ 1000 km$s^{-1}$ \citep{muldrew2012,malavasi2015}.

Here we use the same approach as \cite{darvish2018} . We use a sample of galaxies  within a recessional velocity range of $\Delta$v $=$ $c\Delta$ z $=$ $\pm$ 2000 $kms^{-1}$ corrected by incompleteness due to the flux limit of the sample. The projected surface density is therefore defined as:

\begin{equation}
 \Sigma_{i} = \frac{1}{\psi(D_{i})} \frac{k}{\pi d_{i}^{2}},
\end{equation}
where $\Sigma _{i}$ is the local projected surface density for galaxy $i$, $d_{i}$  is the projected comoving distance to the $k$-th neighbor. ${\psi(D_{i})}$ is the selection function used to correct the sample for the Malmquist bias as a function of the comoving distance according to the following relation:

\begin{equation}
 N(D)dD = AD^{2}\psi (D)dD,
\end{equation}
where
\begin{equation}
 \psi (D) = e^{-\left(\frac{D}{D_{c}}\right)^{\alpha}},
\end{equation}

$A$ is a normalization factor, $D_{c}$ is a characteristic comoving distance corresponding to the peak of the redshift distribution, D is the comoving distance and N(D) is the number of galaxies with measurements of D. The best fit model is given by $A = (2.09 \pm 0.30)\times10^{-3}$, $D_{c} = 1713.39 \pm 106.14 Mpc$, and $\alpha = 3.016 \pm 0.478$ (Figure \ref{select_function1}).

\begin{figure}
    \centering
    \includegraphics[width=.5\textwidth]{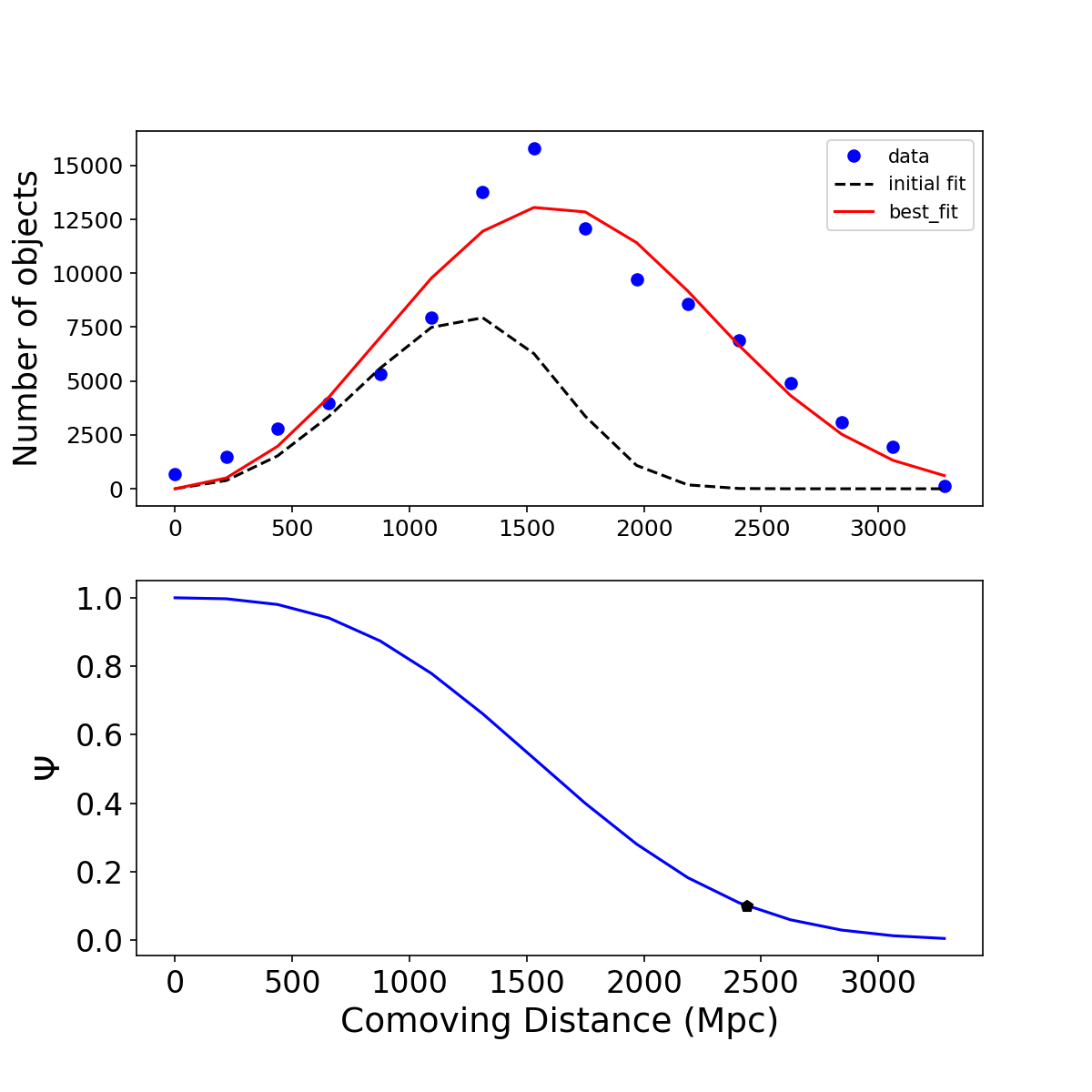}
    \caption{{\it Top}: Comoving distance distribution of the ELBA data (in bins of $\Delta D = 225 Mpc$). The best fit model ($D^{2}e^{-\frac{D}{D_{c}}^{\alpha}}$ ; red solid line) given by $A = (2.09 \pm 0.30)\times10^{-3}$, $D_{c} = 1713.39 \pm 106.14$, $\alpha = 3.016 \pm 0.478$, and  the initial fit (dashed line). {\it Bottom}: Selection function ($\psi$) as function of comoving distance. The black point shows where $\psi = 0.1$. This corresponds to $D = 2441.2660$ Mpc.}
    \label{select_function1}
\end{figure}

Using the selection function we correct the local density measurements for each galaxy by weight $1/ \psi(D)$. The bottom panel in Figure \ref{select_function1} presents the selection function as function of the comoving distance. In this figure the highlighted black point represents the comoving distance value ($D = 2344.4528$) where $\psi (D) =  0.1$. To avoid large uncertainties and fluctuations in the density measurements, we use  galaxies where $\psi (D) \geqslant 0.1$. This corresponds to a comoving distance equals to 2344.4528 Mpc. We did a lower cut of $D = 85$ Mpc that corresponds to z $\sim$ 0.02 to avoid bright sources.

As a sanity check, we attempt to recover the well established relation between galaxy morphologies and their environmental density. We know that in the local universe the fraction of red galaxies depends on stellar mass and environment \citep{baldry2006,peng2010}. Using galaxies down to z $\sim$ 1.0 , \cite{peng2010} showed that the fraction of red and massive galaxies is high for high-density environments. More recently, \cite{kovac2014} have shown that this dependence happens at least out to $z = 0.7$.

Here we define quiescent galaxies as those with specific star formation rates $\log$(sSFR) $\leqslant$ -10.5  $(yr^{-1})$. We calculate environmental densities for all galaxies using the KNN method in the flux-limited sample as described above. We finally determine the fraction of quiescent galaxies in a given density bin, considering the redshift range from $0.08 < z < 0.9$. We note that this redshift range was chosen after the co-moving distance cut was applied, based on the selection function. Figure \ref{red_galaxies_fraction} shows the results of this exercise, and we clearly reproduce the aforementioned relation. We conclude that the method works as expected.

\begin{figure}
    \includegraphics[width=\columnwidth]{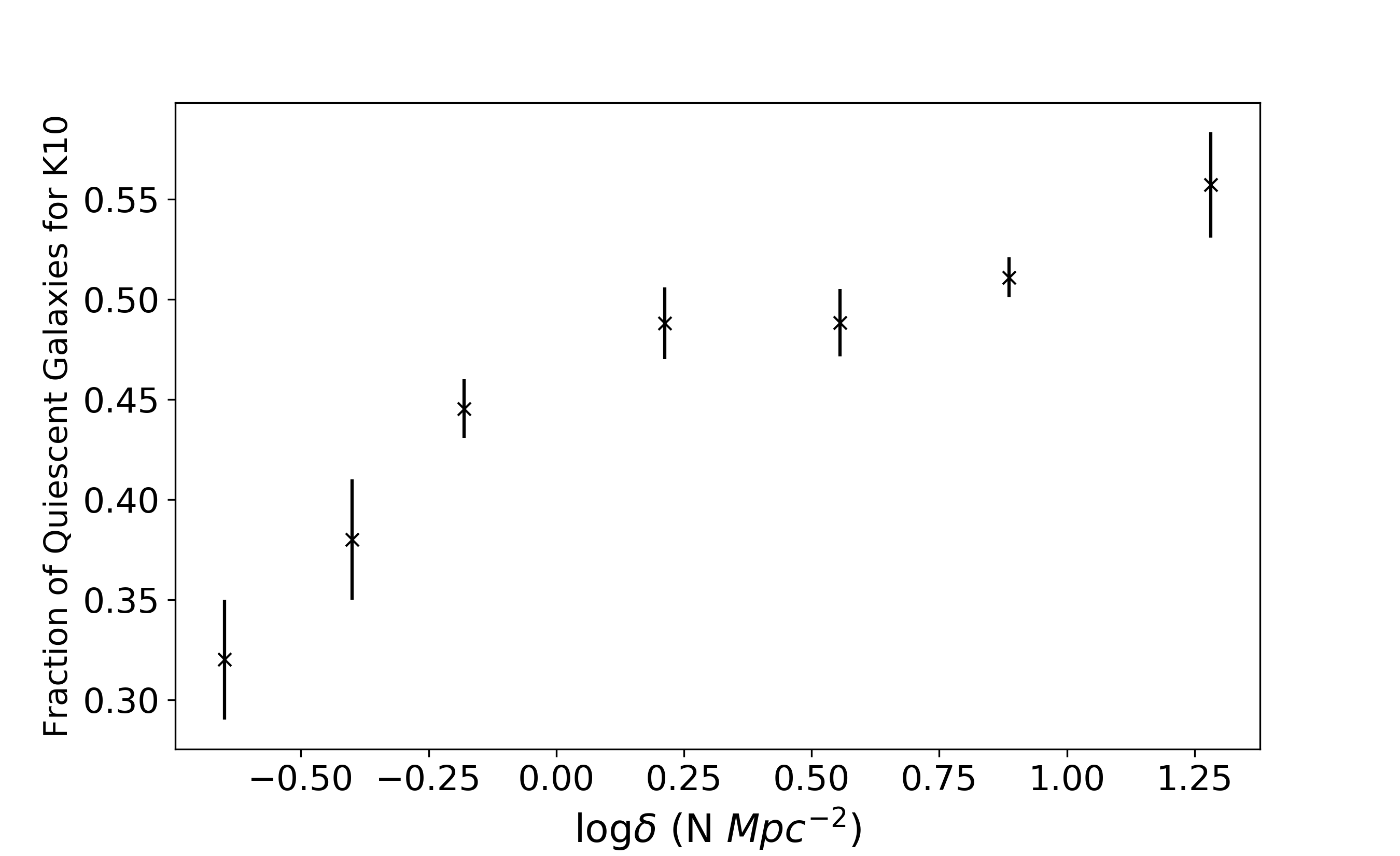}
    \caption{As a test of our local density estimation methodology, we measure the fraction of quiescent galaxies as a function of density (log $\delta$).We notice the clear correlation between specific star formation rates and galaxy environments. In agreement with the morphology-density relation, the fraction of red galaxies is much higher in dense environments.}
    \label{red_galaxies_fraction}
\end{figure}

We remind the reader that the KNN method presents a strong dependence on the number of neighbours (k) used in the analysis, \cite[e.g.][]{darvish2015}. Small values of K (1,2,3) yield unrealistic high-density values because of Poisson noise and clustering of uncorrelated galaxies. Values of k equals to 4, 5 and 10 are usually used \cite[e.g.][]{darvish2018}. $\delta$ values determined for each LBA in the sample are shown in Table \ref{densities_LBA}. We compare density measurements of LBAs and the general galaxy population between 4th, 5th and 10th nearest neighbors (Figure \ref{knn_comp}). The median absolute deviation (MAD) between the density values of the 4th and 5th, 4th and 10th, and 5th and 10th are 0.04, 0.12 and 0.09, respectively (Figure \ref{knn_comp}. Since k5 and k4 presents almost the same values, we will use k4 instead of k5 as a estimator for small scale environment.Our analyses for the sample of LBA will take into consideration the 4th and 10th values obtained using the technique.

\begin{figure}
  \centering
  \includegraphics[width=0.41\textwidth]{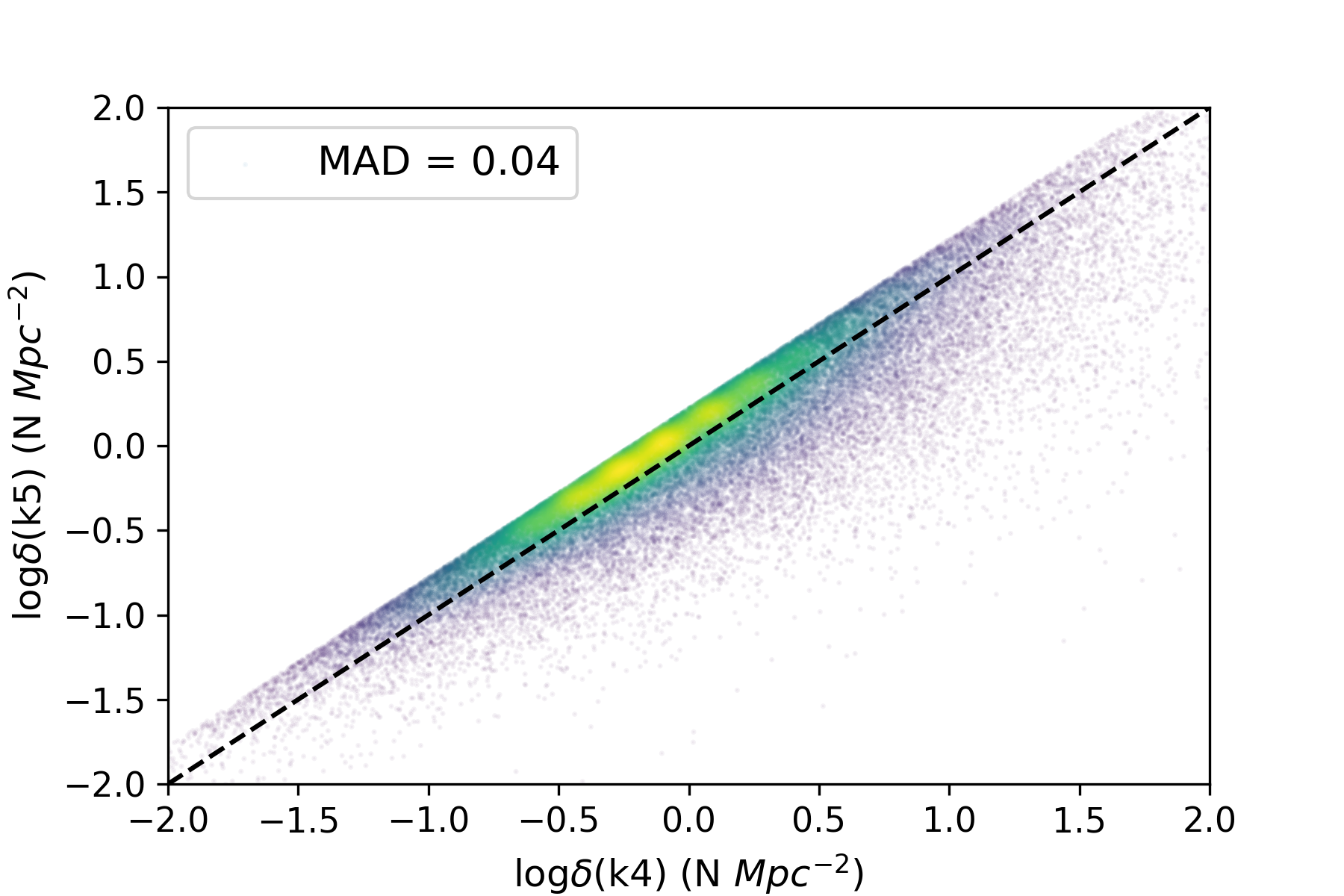}
  \includegraphics[width=0.41\textwidth]{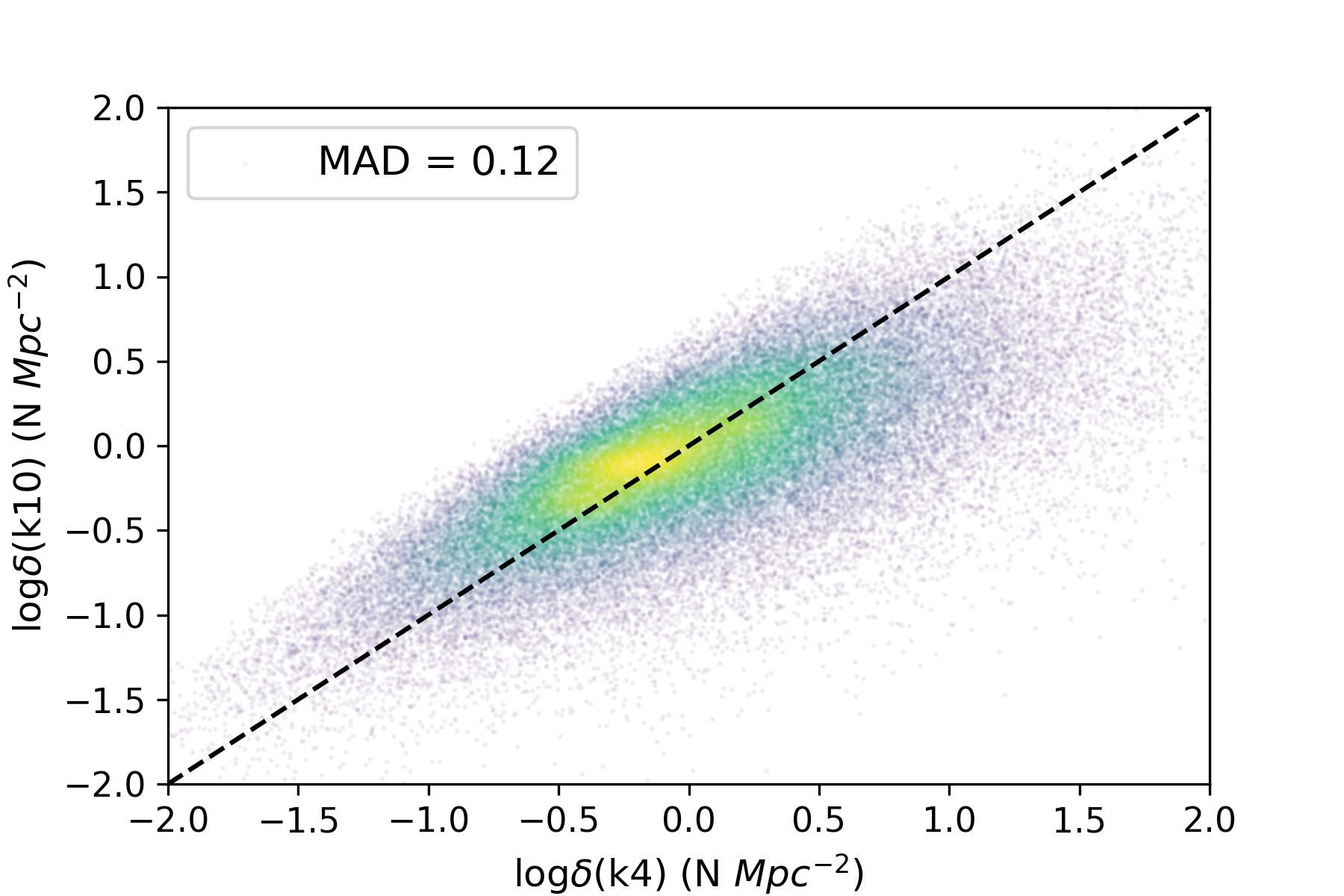}
  \includegraphics[width=0.41\textwidth]{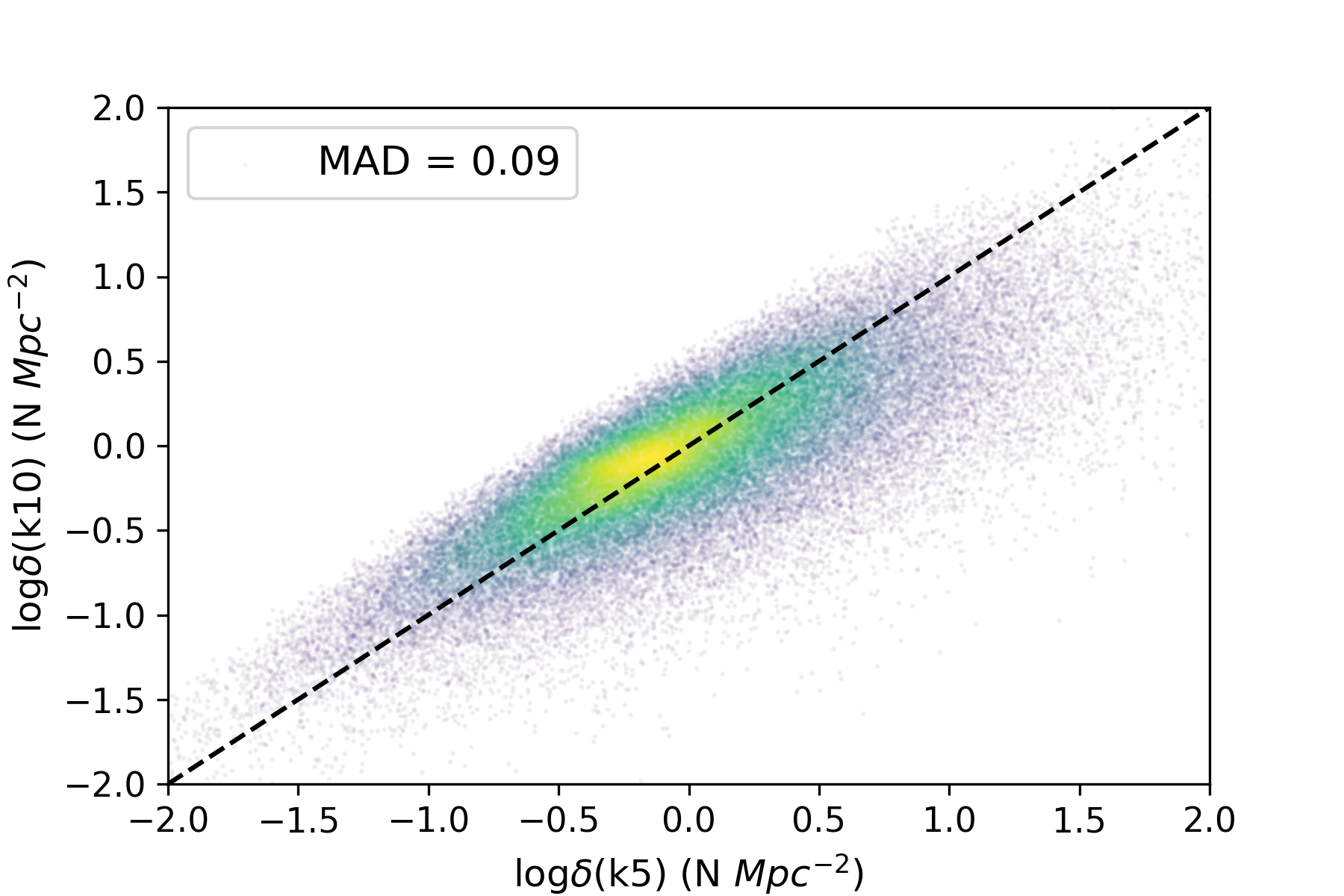}
  \caption{Comparison between the estimated surface density densities using the (top) 4th and 5th, (middle) 4th and 10th, and (below) 5th and 10th nearest-neighbors. The median absolute deviation(MAD) between the density values is also presented in each plot.}
  \label{knn_comp}
\end{figure}

\section{Results and Discussion}
The local environment correlates with many galaxy properties (e.g., star formation history, stellar mass, galaxy morphology, gas kinematics) which are also related to star formation and AGN activity \citep{kauffmann2004,darvish2018, bluck2019}. We note, however, that many of these studies are based on galaxies in the local universe $z \sim 1$. Therefore, the role of environment in the distant universe has not been fully explored, and faces significant observational challenges related to incompleteness in stellar mass and luminosity. Therefore, investigating the local environment of local analogues of high-redshift galaxies offers important constraints on galaxy formation and evolution in the distant universe \cite[e.g.][]{antara2009,adelberger2005,contursi2017,eu2018}.

\subsection{LBAs reside in small galaxy groups or in galaxy pairs}
LBAs were selected based on their compact sizes, replicating properties of high-redshift Lyman break galaxies \citep[e.g.][Figure \ref{ELBA_sample}]{heckman2005,hoopes2007}. Further investigation using higher resolution HST images revealed disturbed morphologies with evidence for mergers or galaxy interactions \citep{overzier2009,overzier2010}. In addition to suggesting recent interactions with nearby neighbors, \citet{overzier2009} showed that similar processes may occur also at $z>3$ LBGs, but that these signatures are less evident due to lower spatial resolution and surface brightness dimming effects.

\cite{antara2009} studied the correlation function of $z\sim 1$ UVLGs and found a higher pair fraction in these galaxies compared to that of other galaxy populations (e.g., late-type or early-type galaxies) in the same field. This suggested that galaxy mergers were the most-likely driver of the the high levels of star formation in these galaxies. Other high-SFR galaxies, such as LIRGS and ULIRGS, also present similarly high pair fractions and assumed merger rates, yet they appear different from ULVGs or LBAs, containing significantly higher quantities of dust.

The aforementioned studies highlight the important role that the small-scale environment plays in high-SFR galaxies. In this work, the main advance in studying the local environment near LBAs results from the application of the nearest neighbor method (see Section \ref{sec:environmentdensity}) on deep images from DECam data. In table \ref{densities_LBA} we present the density measurements obtained using the nearest neighbor estimator for the nine LBAs in our sample. For these galaxies, considering the nearest neighbor value obtained using k10, we found that these galaxies are located within low density regions. Therefore, we conclude the LBAs do not reside in dense galaxy clusters.

Moreover, we find that k4 is generally higher than k10, i.e., the physical density is higher in smaller scales, and therefore-these galaxies may reside within small galaxy groups or pairs. Figure \ref{k4_k10_kstest} shows the density distribution of the general galaxy population, using both the large-scale(k10) and small-scale(k4) estimators. The median values of LBAs is always higher than the median of all galaxies, with the k4 density value being particularly different from the general population.

We perform a Kolmogorov-Sminov (KS) test to verify whether the distributions of density values for the general galaxy population and the LBA sample are statistically distinguishable. We found that, comparing the densities for the estimator k10, the distributions are slightly different. For this case we found a p-value for the null hypothesis(i.e. that they are drawn from the same parent population) of $\sim 30\%$ (statistic=0.3273, p-value=0.29016). However, for the k4 density estimator, the p-value is below $1\%$ (statistic=0.5456, p-value=0.0095), confirming the distinct environments that these galaxy populations inhabit at small scales.

\begin{figure}
   \includegraphics[width=\columnwidth]{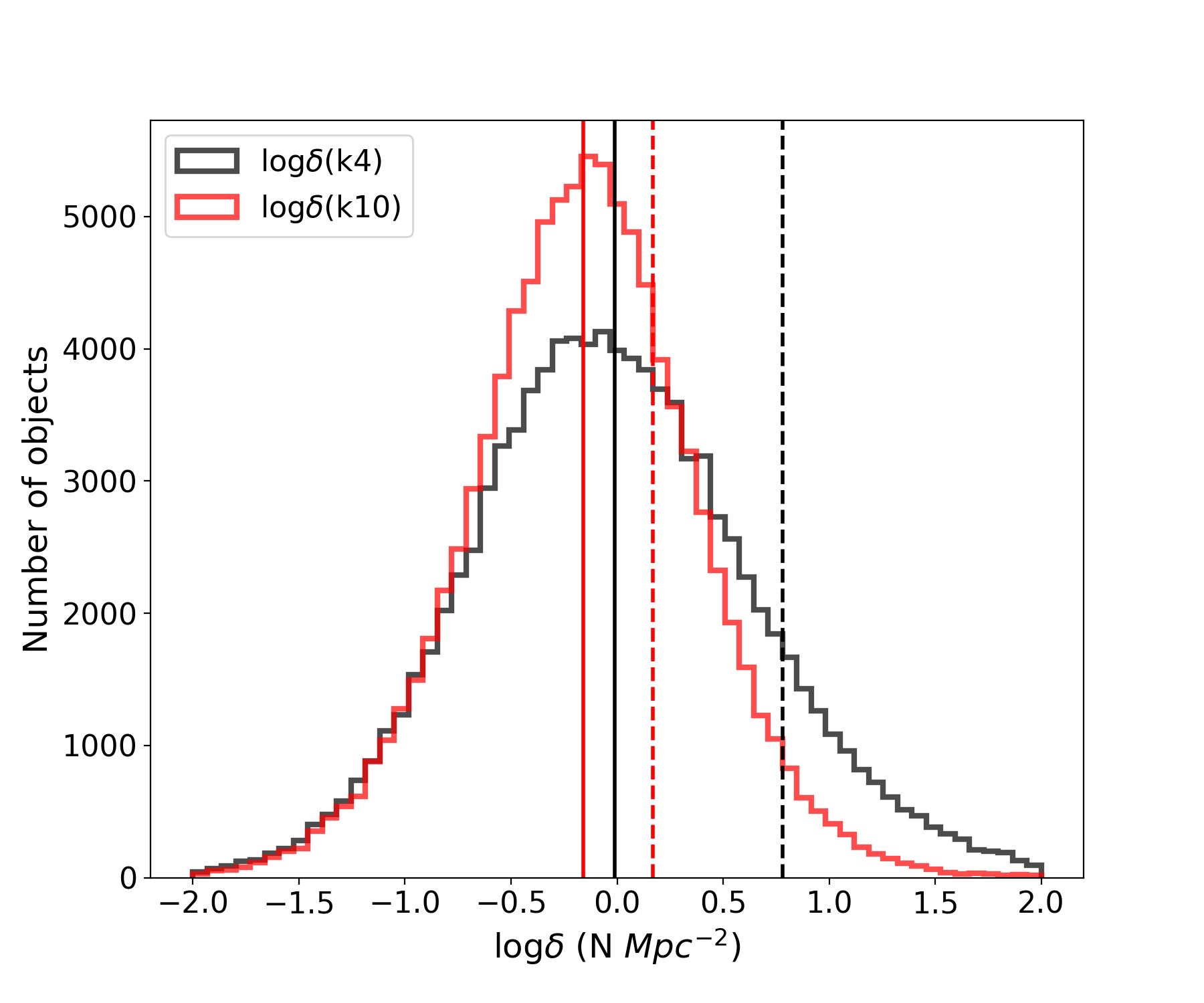}
   \caption{Distribution of the density measurements for the estimators k4(black) and k10(red). Solid lines represents the median value for the measurements obtained for the general population of galaxies, considering the estimators k4(black) and k10(red). Dashed lines represent the median densities just for LBAs, obtained using both estimators(k4, blask dashed line and k10, red dashed line). Comparing the densities for the the estimator k10, the distributions are slightly different. For this case we found a p-value $\sim$ 30$\%$ (statistic=0.3273, p-value=0.29016). However, comparing the densities for the estimator k4, we found that the distributions are completely different.For this case we found a p-value below 1$\%$ (statistic=0.5456, pvalue=0.0095). This measurements, even for a small sample of LBAs, represents a strong evidence that LBAs populated regions denser than the general population of galaxies.}
    \label{k4_k10_kstest}
\end{figure}

Since the merger process can trigger a burst of star formation \cite[e.g.][]{cibinel2019}, we infer that the small-scale environments of these galaxies is relevant to their observed high SFRs. In addition to merger/interaction signatures, the {\it HST} images are useful in identifying clumpy structures, indicative of very massive and concentrated star-forming regions \citep{overzier2009}).
This clumpy star formation is often a result of minor mergers \citep{elmegreen2007,garland2015}. Studying the kinematics of the ionized gas in LBAs, \citet{antara2009b} and \citet{thiago2010} find evidence of dispersion dominated kinematics, consistent with recent merger activity.

We compare the local environment of the LBAs with that of the general population of galaxies in the field by relating the density estimator values with the stellar mass of galaxies (Figure \ref{knn_sfr_mass}).  We divide the general population of galaxies into different stellar mass bins and compare the median stellar mass and density estimator values of the LBAs with those for the general population. To estimate the galaxy properties(stellar masses, SFR), we used LePhare \citep{ilbert2006} with templates from \cite{bruzual2003}. Our comparison allows us to verify that at large scales (k10 estimator, top panel), the value for the LBA sample is consistent with that of the general population, which appears flat over the entire range of stellar mass bins.
However, at small scales (k4 estimator, bottom panel) we observe that the LBAs mean density appears almost 2 dex higher than the general population of galaxies, which suggests that LBAs inhabit environments denser than other galaxies at small scales.
This result and interpretation agree with \cite{antara2009}, who found that UVLGs are mostly found in pairs and small groups. Since LBAs are a subsample of UVLGs with the highest surface brightnesses, we conclude that LBAs occupy small groups of galaxies or pairs, consistent with the prevalence of merger activity inferred  from HST observations \citep{overzier2009} and kinematics studies \citep{antara2009b,thiago2010}. A larger ELBA study containing more LBAs would yield a more statistically robust result.

\begin{figure*}
  \centering
  \includegraphics[width=0.45\textwidth]{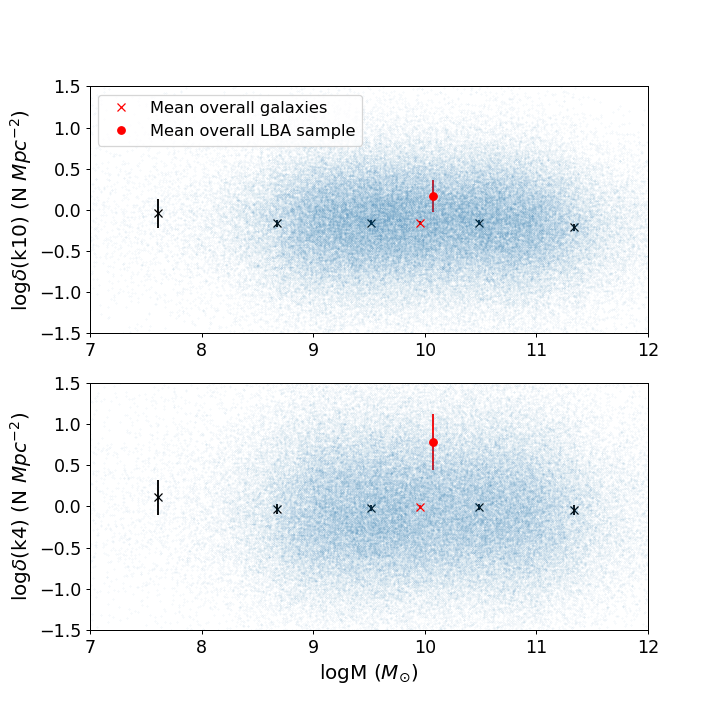}
  \includegraphics[width=0.45\textwidth]{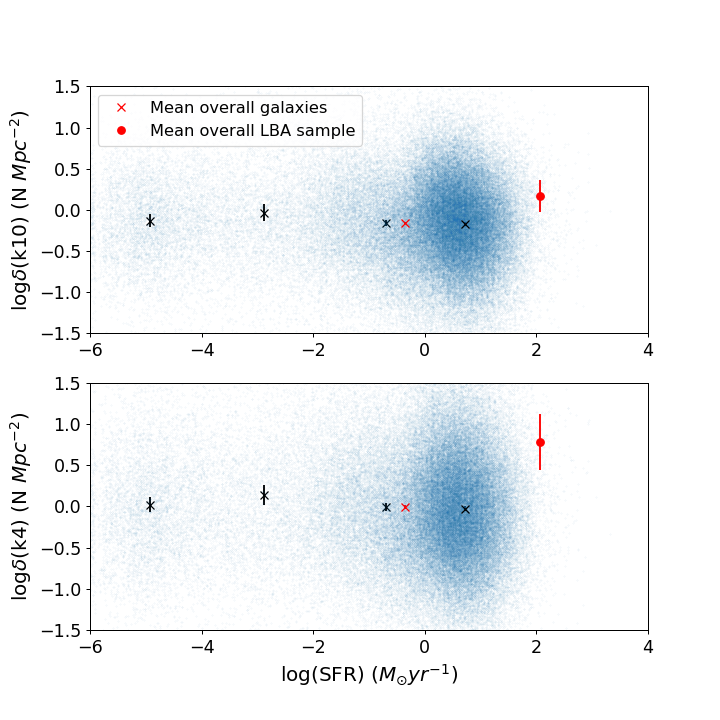}
  \includegraphics[width=0.45\textwidth]{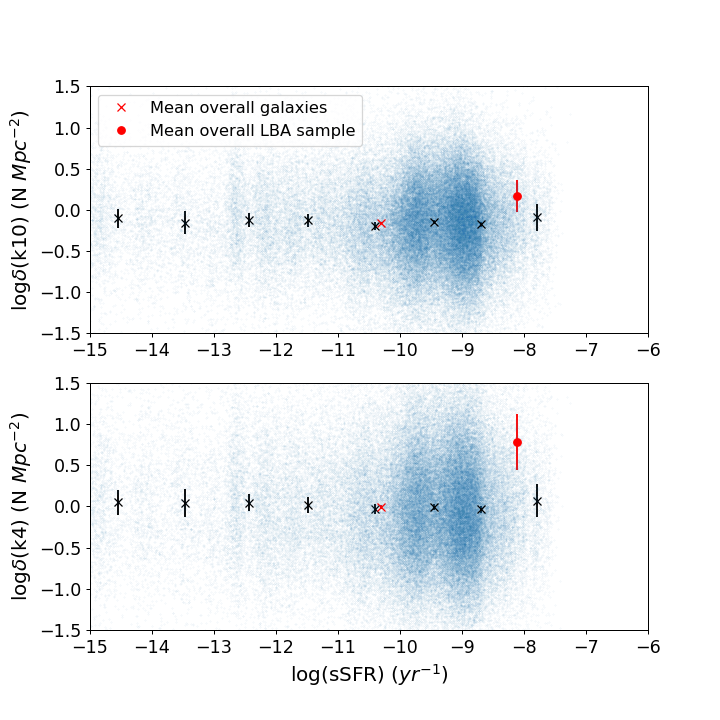}
   \caption{The plots show the relation between each density estimator as function of stellar mass, star formation rate and specific star formation rate. The blue points shows the overall distribution for all galaxies observed on ELBA`s fields. the black x represents the median density for a given stellar mass (top left), star formation rate (top rigth) and specific star formation (bottom) bin with their respective uncertainties. The red x represents the median stellar mass (top left), star formation rate (top right), specific star formation rate (bottom) and median density for the general population of galaxies. The red circle represents the median stellar mass (top left), star formation rate (top right), specific star formation rate (bottom) and density measurement for LBA sample. These plots show that the LBA sample occupies environments denser than the overall populations of galaxies based on small-scale measurements using the 4th nearest neighbour estimator.}
  \label{knn_sfr_mass}
\end{figure*}

\cite{kauffmann2004} suggested a critical stellar mass of $\sim 10^{10.3}$ $M_{\odot}$ as a boundary between two distinct galaxy populations: massive quiescent galaxies and low-mass, highly star-forming objects. LBA stellar masses vary within the range of $9\leq {\rm log} M_{\star}/{M_{\odot}} \leq 11$ \cite[e.g.][]{hoopes2007}.Since LBAs appear similar to high redshift galaxies (e.g., compact morphologies, high specific SFRs, low metallicities and dust attenuations), compared to other nearby (z<1) star-forming galaxies, they present a unique and ideal sample of galaxies for studying galaxy evolution over cosmic time. Many questions remain regarding what drives the high SFRs in these galaxies: mergers due to dense small-scale environments (e.g., close pairs or groups) or increased star formation efficiency caused by gas accretion from the halo \cite[e.g.][]{antara2009}. 

\cite{adelberger2005} used cross-correlation functions to investigate the halo mass in Lyman break galaxies. They found disagreement between the observed and modeled mass for small scales, and argue that this discrepancy is caused by small objects that inhabit the same halo, but are not detected at high redshift due to limitations in sensitivity. In agreement with \cite{adelberger2005} our results suggests that LBAs populate regions denser than average, in comparison with other galaxies in the same field (Figure \ref{knn_sfr_mass}). We argue therefore that the discrepancy found by \cite{adelberger2005} is indeed caused by the non-detection of less massive companions, to which we are more sensitive at low redshift.

\subsection{LBAs as proxies for galaxy preprocessing at high redshift}
The fact that LBAs inhabit groups and small scale overdensities when compared with galaxies of similar stellar mass and SFR could be an evidence of preprocessing. This is reinforced by the observational signatures of tidal effects and mergers, which are also believed to be indicative of preprocessing \citep{mihos2004}.

Galaxy groups are natural structures to observe preprocessing phenomena, which are mainly driven by gravitational interactions \cite[e.g.][]{mihos2004}. Part of the expected transformations in galaxy structure happen in early conditions that are not easily observed in galaxy clusters, but can happen in small scale structures such as galaxy pairs and in filaments \citep{dressler2004}.  Furthermore, we estimate that at least 50\% of the LBAs are classified as mergers due to disturbed kinematics or morphologies \citep{thiago2010,overzier2010}. Therefore, we argue that the small-scale environment does play a role in shaping these galaxies and regulating star formation.

At the same time, assuming the analogy with the high-redshift universe is valid, both LBAs and LBGs are main-sequence star-forming galaxies, i.e. their star formation rates are similar to the typical star-forming galaxies of same stellar mass at that redshift ($z\sim 1.5-2.0$ in the case of LBAs)\citep{contursi2017}. That would indicate that the merging process is not violent enough to produce strong starburst episodes that will subsequently rapidly quench star formation. Instead, we argue that the observed processes are consistent with a picture in which gas inflows from the intergalactic medium and "wet" (gas-rich) mergers fuel a nearly steady state, in which violent disk instabilities promote compaction and bulge growth at high redshift. This is consistent with several predictions from simulations and actual observations of the distant universe \citep{barro2013,ceverino2015}.  Indeed, we do observe the bulge growth in several objects in the LBA sample. \cite{overzier2009}  have shown that $6$ out of $30$ objects observed with the HST, or $20$ percent of the sample, have DCOs whose bolometric luminosities are dominated not by AGN activity but by star formation \citep{jia2011,alexandroff2012}, within very dense regions smaller than  $100 pc$. These could be the result of the loss and transference of angular momentum necessary to transport material to the central regions of the galaxy, leading to bulge growth and subsequent decrease of SFRs.

%Indeed, we do observe the bulge growth in several objects in the LBA sample. \cite{overzier2009} have shown that $6$ out of $30$ objects observed with the HST, or $20$ percent of the sample, have DCOs whose bolometric luminosities are dominated not by AGN activity but by star formation \citep{jia2011,alexandroff2012}, within very dense regions smaller than  $100 pc$. These could be the result of the loss and transference of angular momentum necessary to transport material to the central regions of the galaxy, leading to bulge growth and subsequent decrease of SFRs.
%%%TeX ver um problema nesse paragrafo acima

\begin{table}
    \centering
    \caption{Density measurements obtained using the KNN method for each LBA in the sample. All the values was corrected by the selection function.}
    \label{densities_LBA}
    \begin{tabular}{c c c c }
    \hline
    \hline
    Galaxy & $\log \delta _{k4}$               & $\log \delta _{k5}$                & $\log \delta _{k10}$  \\
    LBA223& -0.31            & -0.75             & -0.49              \\
    LBA231& 0.78             & 0.21              &  0.13       \\ 
    LBA218& 1.05             &1.38               &0.55 \\
    LBA238& 1.66             &1.55               &1.18       \\
    LBA242& 0.04             &0.16               &-0.39      \\
    LBA246&2.04              &1.01               &0.45        \\
    LBA326&1.44              &1.22               &0.74          \\
    LBA334&1.57              &1.02               &-0.58         \\
    LBA349&-1.25             &-1.07              &-0.08         \\
    
    \hline
    \hline
    \end{tabular}
\end{table}

\section{Summary and Conclusions}

In this paper, we study the environment of LBAs using the photometric data from our ELBA survey. LBAs share similar properties with high-redshift star-forming galaxies and work as excellent proxies for the study of galaxy formation in the distant universe. To measure the local environment for each galaxy in the sample we apply the KNN method following the methodology of  \citet{darvish2018}. Our main contributions are the following.

\begin{itemize}
 \item The ELBA survey provides deep photometric data $\sim$2 mags deeper than SDSS in the g band for 10$\sigma$ point source detection for 33 square degrees in the sky. This is 1 magnitude deeper than DES in the same band. This allows for an investigation of the environments occupied by different galaxy populations down to fainter limits, probing less massive galaxies. This data will be made publicly available and can be used by the community at large for a number of galaxy formation and evolution studies.

\item The photo-z measurements in this work , using u,g,r and i bands from DECam and J,H,K from UKIRT \citep{ukidss}, are consistent with other papers that use several (ie 4--10) photometric bands \cite[e.g.][]{sanchez2014}. We use these redshifts to determine the environmental densities of all objects in our catalogue.

\item Based on the density estimator k4 we conclude that LBAs, reside in small groups (pairs or systems of up to 4 members), as opposed to massive galaxy clusters. The median density value for LBAs is similar to that of the general population of galaxies. This is based on density estimators for larger structures (namely the distance to the 10-th nearest neighbour, k10).

\item Analysing density estimators for smaller scale structures ($k4$) we find that the median density of LBAs is higher than that of other galaxies of similar stellar mass, star formation rates and specific star formation rates. We interpret this as evidence for small scale clustering in LBA environments. This result agrees with merger signatures observed in high-resolution images \cite[e.g.][]{overzier2009} and previous studies that find LBAs to have high pair fractions \cite[e.g.][]{antara2009}. These conclusions are also consistent with results for star-forming galaxies at high redshift \citep{adelberger2005}.

\item We interpret our results as indication that LBAs are a sample of galaxies located in low redshift that are passing through early stages of preprocessing. This process, both for LBAs and high-redshift star-forming galaxies, does not necessarily generate strong starburst events, but lead to disk instabilities and compaction that will yield bulge growth and the gradual quenching of star formation.

\end{itemize}

Future work will apply the same analysis for other high-z analogues samples such as emission line galaxies and green peas \cite[e.g.][]{sangeeta2012,greenpeas,yang2017}. This will enable a similar investigation of the influence of environment on the properties of distinct galaxies, that might represent good low-redshift laboratories of earlier cosmological epochs. We will use our observations to constrain the predictions from cosmological simulations \cite[e.g. the IllustrisTNG;][]{pillepich2019} and better understand the physics that drives higher SFRs in LBAs and LBGs.

\section*{Acknowledgements}

We thank Behnam Darvish for the help during our density measurenments tests. GAZPAR team.The Cerro Tololo Inter-American Observatory (CTIO) by the instrumentation(Blanco and DECam) used in this work. CAPES by  scholarship.

%% The best way to enter references is to use BibTeX:

\bibliographystyle{mnras}
\bibliography{elba}

%\bibliographystyle{mnras}


%%%%%%%%%%%%%%%%%%%%%%%%%%%%%%%%%%%%%%%%%%%%%%%%%%

%%%%%%%%%%%%%%%%% APPENDICES %%%%%%%%%%%%%%%%%%%%%

%\appendix

%%%%%%%%%%%%%%%%%%%%%%%%%%%%%%%%%%%%%%%%%%%%%%%%%%

% Don't change these lines
\bsp	% typesetting comment
\label{lastpage}
\end{document}